\documentclass[a4paper]{amsart}
\usepackage{amsmath, amsthm,amssymb}

\usepackage{todonotes}

\theoremstyle{definition}

\theoremstyle{remark}

\title{IGLOSS: iterative gapless local similarity search }

\author[Rabar \textit{et~al}.]{Braslav Rabar$^{\text{1}}$, 
	Strahil Ristov$^{\text{2}}$, 
	Maja Zagor\v{s}\v{c}ak$^{\text{3}}$, 
	Martin Rosenzweig$^{\text{1}}$, \\
	Pavle Goldstein$^{\text{1}}$
	} 

\address{$^{\text{\sf 1}}$Mathematics Department, Faculty of Natural Sciences and Mathematics, Bijeni\v{c}ka 30, 10000 Zagreb, Croatia
	$^{\text{\sf 2}}$Ru\dj{}er Bo\v{s}kovi\'{c} Institute, Bijeni\v{c}ka 54, 10000 Zagreb, Croatia
	$^{\text{\sf 3}}$Department of Biotechnology and Systems Biology, National Institute of Biology, \\ Ve\v{c}na pot 111, 1000 Ljubljana, Slovenia
}

\email{pavle.goldstein@math.hr}

\begin{document}

	\begin{abstract}
	 Searching for local sequence patterns is one of the basic tasks in bioinformatics. Sequence patterns might have structural, functional or some other relevance, and numerous methods have been developed to detect and analyze them. These methods often depend on the wealth of information already collected. The explosion in the number of newly available sequences calls for novel methods to explore local sequence similarity. We have developed a high sensitivity web-based iterative local similarity scanner, that finds sequence patterns similar to a submitted query.\\ 
		\textbf{Availability:} The IGLOSS web server is available at {http://compbioserv.math.hr/igloss/}\\
	\end{abstract}
	
	\maketitle

\section{Introduction}

Motif scanning methods are at the heart of many bioinformatics procedures. 
For example, secondary structure recognition, 
proteome annotation \cite{Zhang2003} and, in general, protein family assignment \cite{Finn2007}, all depend - to a certain extent - on detecting a variant of a (amino acid) motif in a given sequence or a set of sequences. 
Consequently, numerous methods - Smith-Waterman algorithm \cite{Waterman1976}, BLAST \cite{Altschul1990}, PSSM \cite{Gribskov1987}, Viterbi \cite{Viterbi1967} - and applications - e.g. FIMO \cite{Grant2011} - have been developed. 
Typically, the application takes a motif profile as an input, and then, using a dynamic programming approach, or some approximation, finds a version of optimal local alignment in each scanned sequence. 
Clearly, accuracy of this approach depends - among other things - on the motif profile being of a sufficient quality and size. 
A considerable increase in the number of newly available sequences, where only a small portion has been properly analysed, makes the task of creating an unbiased, representative profile increasingly difficult, and necessitates a different approach. 

In this note, we present IGLOSS -- iterative gapless local similarity search -- a web-application that will, in a proteome or a collection of proteomes, find sequence patterns similar to a submitted query. 
The query can consist of one or more sequences of equal length, the level of required similarity can be easily controlled, and we provide simple options for conserved/neutral positions, as well.

\vspace*{0pt}
\section{Implementation}

Our iterative approach is implemented in a straightforward fashion: initially, a crude profile - with crudeness depending on the size of the query - is built, and the dataset is scanned, with the maximal log-odds score reported for each sequence. 
Standard results (see \cite{DeHaan2006}) guarantee that the scores will be approximately logistically distributed, and motifs with scores above the predetermined scale are used to build a new profile. 
Clearly, this procedure stops when there is no change in the list of positives, or the predetermined number of iterations is reached. 

Let us assume that we are given a motif $M$, of length $k$. To maximize log-odds scores, we compute, for each sequence $x$, the log-odds vector $v(x)$. 
The components of $v(x) = (v(x)_1, v(x)_2, \ldots , v(x)_{n-k+1})$ are given by   
\begin{equation}
v(x)_l = \sum_{i=0}^{k-1} \log \frac{ P(x_{l+i} | y_{i+1} ) }{ P( x_{l+i} | q )}, \ \ l=1, \ldots, n-k+1
\end{equation}
where $ \{ y_{1}, \ldots , y_{k} \} $ and $q$ are distributions determining the position specific scoring matrix for $M$ (in other words, a motif profile). 
The vector $v(x)$ is computed using a modification of the fast indexed string matching algorithm from \cite{Ristov2016}, which considerably reduces overall processing time. Distributions $ \{ y_{1}, \ldots , y_{k} \} $ - or, rather, $ \{ y_{1}^{(j)}, \ldots , y_{k}^{(j)} \} $ - represent emission probabilities for each column of $M$ in $j$-th iteration, while $q$ is the background distribution for the standard amino acid alphabet. 
Clearly, $ \{ y_{1}^{(j)}, \ldots , y_{k}^{(j)} \} $ - and the way they are refined through iterations - are among the essential aspects of the iteration process. 
We compute these distributions from the list of positives from the previous iteration - or just the query, for the first iteration. 
We use a mild sequence-weighting scheme (from \cite{Henikoff1994}), and, depending on the query size, add evolutionary pseudo-counts, given by a high power of the PAM matrix \cite{Dayhoff1978}. Finally, we average this with the initial distributions -  denoted by $ \{ y_{1}^{(1)}, \ldots , y_{k}^{(1)} \} $ - in order to prevent divergence. 
We give a detailed description of this procedure on the server website.

\vspace*{0pt}
\section{Example and Evaluation}
We applied our server to the GDSL-lipase protein family from four higher plants - {\em Arabidopsis thaliana} (AT) (TAIR9), {\em Oryza sativa} (OS) (MSU v7), {\em Solanum tuberosum} (ST)  (ITAG1), and {\em Solanum lycopersicum} (SL) (ITAG2.3).
Proteins in this family display fairly low overall sequence similarity, but are reasonably well described by the presence of five characteristic motifs (also called blocks), with blocks I, III and V being more conserved  (\cite{Chepyshko2012a}, \cite{Vujaklija2016}). 
The evaluation consisted of submitting a single sequence, typical for block I, to our scanner, and checking the annotation of sequences in which positive hits are found. 
We measure the quality of our results by computing {\em true positive rate} (TPR), i.e. {\em sensitivity}, and {\em positive predictive value} (PPV), i.e. {\em precision}, and compare them to those obtained by iterative BLAST (i.e. PSI-BLAST, \cite{Altschul1997}). 
PSI-BLAST was used with the same input and with the standard parameters, and with e-value adjusted to obtain approximately the same number of positives.  
Annotation was determined by further processing the information from GoMapMan \cite{Ramsak2014a} resource. The sizes of the resulting protein families (per organism) are given in the central column of Table~\ref{Tab:01}.
All data sets -- proteomes, lists of positives, as well as queries and a detailed description of the processing procedure - are available from the server website.

\begin{table}[!h]

	\centering
	\begin{tabular}{|cccccc|c|ccccc|}
		\hline
		\rule[-1ex]{0pt}{5.0ex} \tiny Organism  & \tiny  P   &  \tiny TP  &  \tiny PPV &  \tiny TPR &  \tiny scale  & \tiny condition positive  & \tiny blast P & \tiny blast TP & \tiny blast PPV & \tiny blast TPR & \tiny evalue \\
		\hline \hline 
		\rule[-1ex]{0pt}{5.0ex} AT              & $106$      & $85$       &  0.80      &  0.72      & 8             &  $118(104)$          &  106          &  75            &  0.71           &    0.64         & 480 \\ 
		\hline 
		\rule[-1ex]{0pt}{5.0ex} OS              & $114$      & $108$      &  0.95      &  0.70      &  10           &  $155(116)$          &  115          & 86             &  0.75           &    0.55         & 900 \\ 
		\hline 
		\rule[-1ex]{0pt}{5.0ex} ST              & $124$      & $89$       &  0.72      &  0.72      & 7.5           &  $123(123)$          &  123          &  76            &  0.62           &    0.62         & 600 \\ 
		\hline 
		\rule[-1ex]{0pt}{5.0ex} SL              & $108$      & $89$       &  0.82      &  0.82      & 8             &  $108(108)$          &  107          & 76             & 0.71            &    0.70         & 470 \\ 
		\hline 	
	\end{tabular}
	\caption{ Results for GDSL-like motif query FVFGDSLSDA. The results for IGLOSS
		are presented on the left and the results for PSI-BLAST on the right. Labels: P -- positive, TP -- true positive, CP -- condition positive, PPV=TP/P, TPR=TP/CP; in brackets is the size of the family (per organism), with isoforms counted once }
	\label{Tab:01}
\end{table}

As can be seen from Table~\ref{Tab:01}, our application is more accurate than PSI-BLAST, with our PPV and TPR consistently higher.
While PSI-BLAST is a more versatile and much faster application - GDSL-lipase search on a rice proteome takes around 60 seconds with IGLOSS, and a couple of seconds with PSI-BLAST - this decrease in speed is accompanied by a significant increase in accuracy. Another option for a comparison would be the HMMer-suite, and its iterative module \cite{Finn2015}, but we selected PSI-BLAST, primarily for the ease of use. It should also be pointed out that a comparison with non-iterative methods would be unfair. The mathematical background of our method is very similar to that of BLAST, with main differences in implementation being purpose-specific model building and the distribution parameter estimation. While this is certainly time consuming - especially the latter - it appears that it contributes towards considerably greater accuracy. 

In conclusion, we suggest that our method is a viable alternative when it comes to motif scanning, protein family analysis, and even proteome annotation. On the other hand, while IGLOSS can be used as a fast iterative motif scanner, its primary aim is to help researchers explore common sequence patterns in a proteome.

\section*{Funding}

This work has been partially supported by the European Regional Development Fund [KK.01.1.1.01.0009 - DATACROSS] (SR), the Slovenian Research Agency [J7-7303] (MZ) and
the Croatian Science Foundation [IP-2014-09-2285] (BR) and [IP-2016-06-1046] (PG).

\end{document}